\documentclass[conference]{IEEEtran}
\IEEEoverridecommandlockouts
\usepackage{cite}
\usepackage{amsmath,amssymb,amsfonts}
\usepackage{algorithmic}
\usepackage{graphicx}
\usepackage{textcomp}
\usepackage{xcolor}
\usepackage{multirow}
\def\BibTeX{{\rm B\kern-.05em{\sc i\kern-.025em b}\kern-.08em
    T\kern-.1667em\lower.7ex\hbox{E}\kern-.125emX}}
\begin{document}

\title{A 97 fJ/Conversion Neuron-ADC with Reconfigurable Sampling and Static Power Reduction \\
	\vspace{-0.5cm}
	\author{\IEEEauthorblockN{Jinbo Chen$^{1,2}$, Hui Wu$^{2}$, Jie Yang$^{2}$, and Mohamad Sawan$^{2}$, \IEEEmembership{Fellow, IEEE}}
		\IEEEauthorblockA{$^{1}$Zhejiang University, Hangzhou, Zhejiang, China 310058 \\
			$^{2}$CenBRAIN Neurotech, School of Engineering, Westlake University, Hangzhou, Zhejiang, China 310024 \\
			\ Email: yangjie@westlake.edu.cn, sawan@westlake.edu.cn}
		\vspace{-1cm}
	}
}
\maketitle
	

%
%
%


\begin{abstract}
A bio-inspired Neuron-ADC with reconfigurable sampling and static power reduction for biomedical applications is proposed in this work. The Neuron-ADC leverages level-crossing sampling and a bio-inspired refractory circuit to compressively converts bio-signal to digital spikes and information-of-interest. The proposed design can not only avoid dissipating ADC energy on unnecessary data but also achieve reconfigurable sampling, making it appropriate for  either low power operation or high accuracy conversion when dealing with various kinds of bio-signals. Moreover, the proposed dynamic comparator can reduce static power up to 41.1\% when tested with a 10 kHz sinusoidal input. Simulation results of 40 nm CMOS process show that the Neuron-ADC achieves a maximum ENOB of 6.9 bits with a corresponding FoM of 97 fJ/conversion under 0.6 V supply voltage.

\end{abstract}


\section{Introduction}

With the development of biomedical technology and aging society, recent years have witnessed the rapid advance of bio-signal recording systems for personal healthcare. The conventional architecture of bio-signal recording system includes a pre-amplification stage. Then, the input signal is filtered and digitalized by SAR ADC, and sent to a wireless transmission module or signal processing engine for disease detection \cite{9766042}. Limited by the power budget of batteries and wireless power transmission, energy efficiency is critical to such systems. However, conventional system architectures adopt SAR ADC and Nyquist sampling, suffering from critical energy consumption issues. The highest frequency band of bio-signals determines the sampling frequency, which does not leverage bio-signals features and sparsity. A sparse bio-signal is repetitively sampled even during the silent signal period, thus wasting ADC power on needless data. 

\begin{figure}[t]
	\centerline{\includegraphics[width=3in]{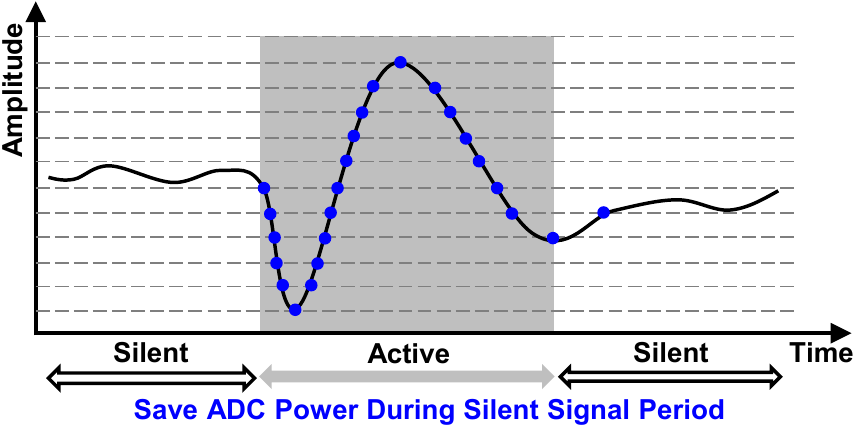}}
	\vspace{-0.3cm}
	\caption{Illustration of level-crossing sampling \cite{9766042}.}
	\vspace{-0.4cm}
	\label{fig_sampling_scheme}
\end{figure}

Recently, level-crossing ADC (LC-ADC) has been developed as a promising candidate for biomedical applications \cite{9789561}. It leverages signal sparsity and achieves event-driven sampling with lower average sampling rate. As presented in Fig. \ref{fig_sampling_scheme}, the input signal is only sampled when predefined levels are crossed, thus saving ADC energy during the silent signal period and realizing data compression compared with Nyquist sampling. Moreover, LC-ADC is based on non-uniform sampling and time quantization, so it produces no aliasing at the output spectrum and is more suitable for low operation voltage and advanced nodes. Additionally, outputs of level-crossing sampling are inherently compatible with neuromorphic spiking neural networks, paving the way to neuromorphic recording and processing systems \cite{chen2022event}.

\begin{figure}[t!]
	\centerline{\includegraphics[width=3.5in]{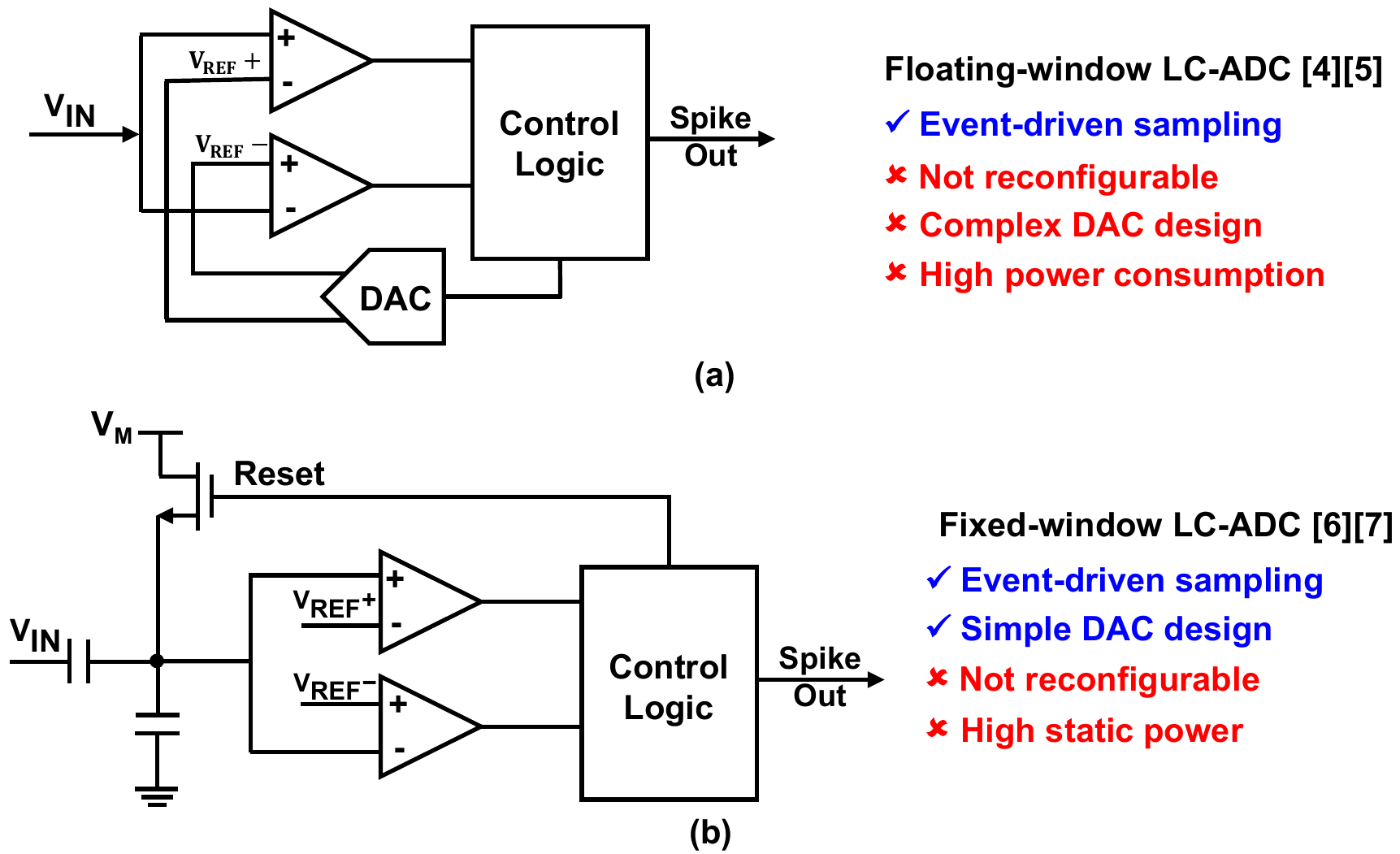}}
	\vspace{-0.3cm}
	\caption{Conventional architectures of LC-ADC. (a) Fixed-window LC-ADC. (b) Floating-window LC-ADC.}
	\vspace{-0.6cm}
	\label{fig_prior_system}
\end{figure}

\begin{figure}[t!]
	\centerline{\includegraphics[width=3.5in]{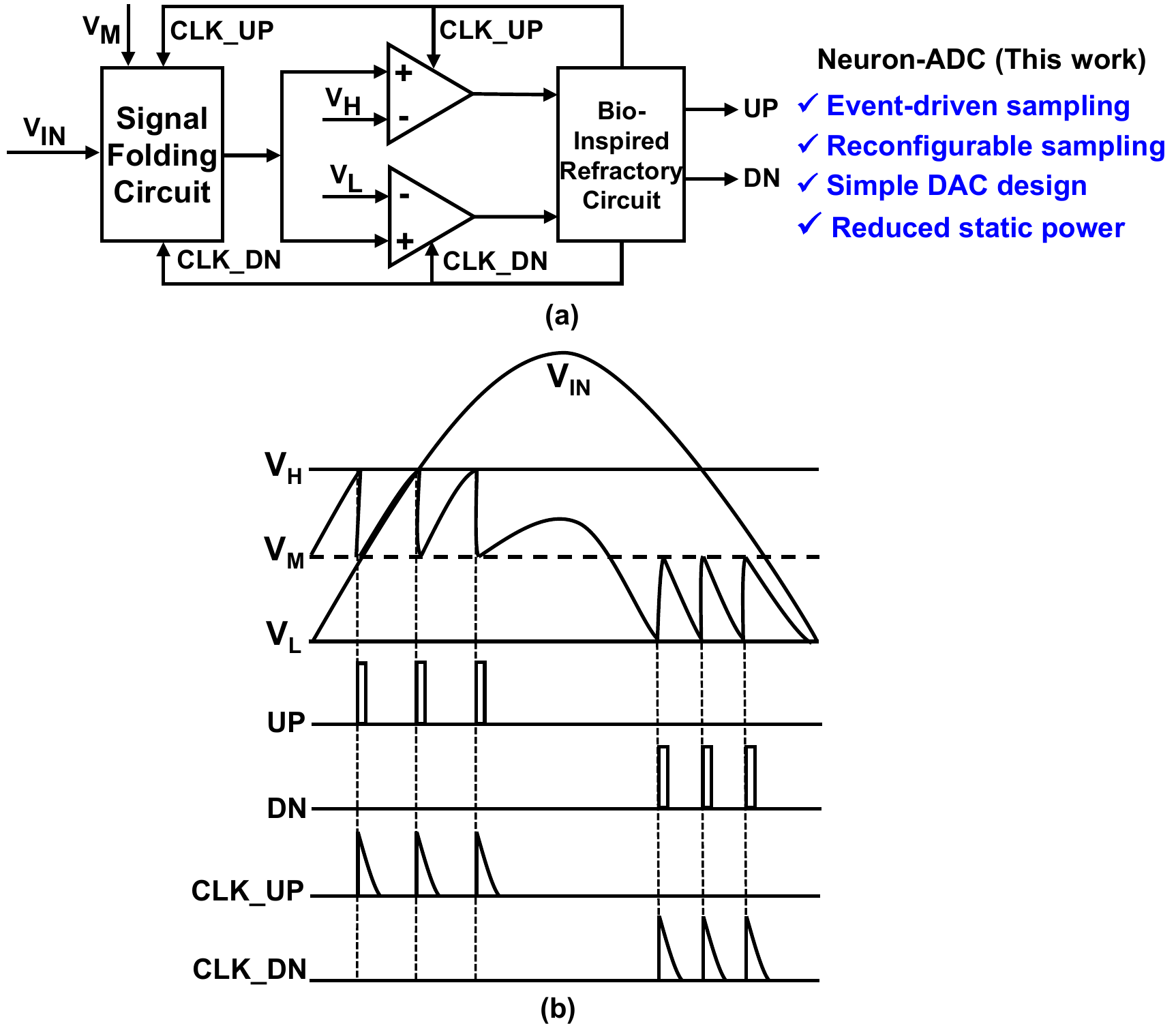}}
	\vspace{-0.3cm}
	\caption{The proposed Neuron-ADC. (a) System architecture. (b) Example operation waveforms.}
	\vspace{-0.6cm}
	\label{fig_proposed_system}
\end{figure}

Prior architectures of LC-ADC suffer from several design drawbacks and performance issues. As shown in Fig. \ref{fig_prior_system}(a), the floating-window LC-ADC proposed in \cite{schell2008continuous}\cite{zhang2014300} is composed of an n-bit DAC, two comparators, and a control logic module. This type of architecture needs complex DAC design that consumes high power. Moreover, circuit parameters and output data rate of this architecture are not reconfigurable and cannot be adapted to various bio-signals. The fixed-window LC-ADC illustrated in Fig. \ref{fig_prior_system}(b) reduces the energy consumption by utilizing a simple 1-bit capacitive DAC that generates a fixed range input to comparators \cite{hou20181}\cite{yazdani2021low}. However, the sampling scheme and output data rate are still not reconfigurable. What's worse, high static power of comparators dominates the power budget and is independent of input signal activity, which counteracts advantages of event-driven sampling.

To solve the aforementioned issues, this work proposes a Neuron-ADC with reconfigurable sampling and static power reduction for biomedical applications. The proposed Neuron-ADC receives inputs of various bio-signals, and produces digital spike outputs with refractory period similar to biological neurons. As shown in Fig. \ref{fig_proposed_system}(a), it is based on the fixed-window LC-ADC architecture. The main contributions are listed as follows:
\begin{itemize}
	\item Propose a bio-inspired refractory circuit to achieve reconfigurable sampling.
	\item A charge-injection-free signal folding circuit is designed to perform as a 1-bit DAC.
	\item Dynamic comparators are utilized to reduce static power consumption.
	\item Simulation results under 40 nm CMOS process show that the Neuron-ADC achieves the ENOB up to 6.9 bits with a corresponding FoM of 97 fJ/conversion under 0.6 V supply voltage.
\end{itemize}

The rest of this paper is organized as follows. Section II introduces the system architecture of the proposed Neuron-ADC, followed by the circuit implementation in Section III. Section IV presents the simulation results. Finally, Section V concludes the paper.


\section{System Architecture}

The system architecture of the proposed Neuron-ADC is shown in Fig. \ref{fig_proposed_system}(a). It includes a signal folding circuit, two dynamic comparators, and a bio-inspired refractory circuit. As illustrated in Fig. \ref{fig_proposed_system}(b), no output is generated when input signal is between two levels of the fixed window, $V_{H}$ and $V_{L}$. When signal crossing happens, the input signal will be converted to two types of digital spike outputs, UP and DN, to represent up crossing and down crossing, respectively. Specifically, when the input signal passes through the predefined levels, the output of one comparator goes high and activates the refractory circuit. The refractory circuit generates two types of outputs: UP/DN, and CLK\_UP/CLK\_DN. The former type of output can be connected with continuous-time neuromorphic processing systems. The latter is with adjustable refractory period and directly fed back to the signal folding circuit and dynamic comparators, performing as control signals to achieve reconfigurable sampling and static power reduction. 

In the proposed Neuron-ADC, the signal folding circuit resets input signals to $V_M$ when receiving CLK\_UP/CLK\_DN signals. The refractory period determines the maximum operation frequency of the signal folding circuit, and output data rate of the Neuron-ADC, thus configuring the output data sampling and compression. During the signal folding process, no signal comparison is needed. Therefore, the two dynamic comparators will be turned off with the time length defined by the refractory period. In this way, the static power is reduced compared with prior works using static comparators.

\section{Circuit Implementation}

\subsection{Charge-Injection-Free Signal Folding Circuit}
\begin{figure}[t]
	\centerline{\includegraphics[width=3.5in]{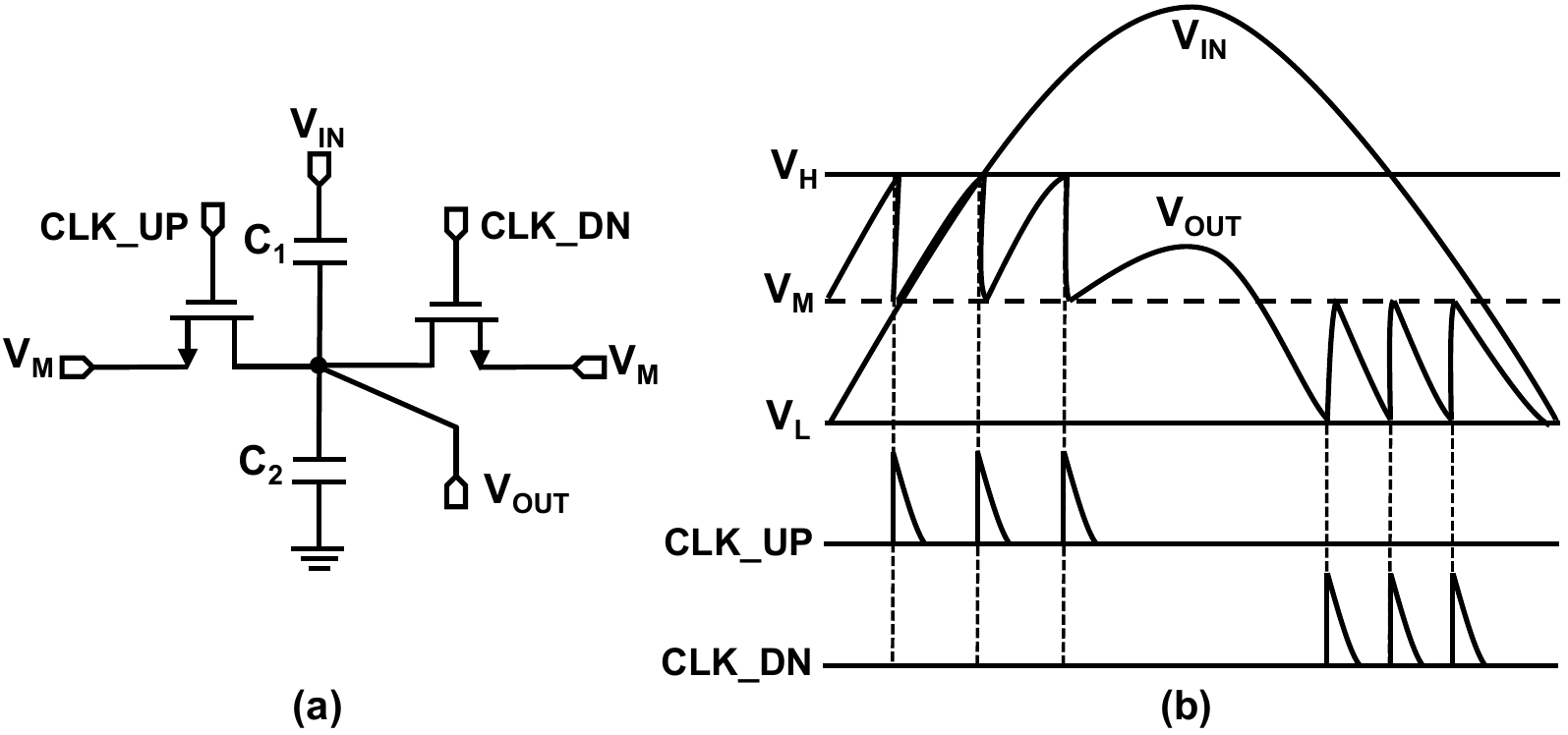}}
	\vspace{-0.3cm}
	\caption{The proposed signal folding circuit. (a) Circuit schematic. (b) Timing diagram.}
	\vspace{-0.7cm}
	\label{fig_signal_folding}
\end{figure}

The signal folding circuit receives input analog signal and asynchronous clock signal from the bio-inspired refractory circuit. It produces a fixed range of output analog signal to be compared with reference voltage levels by the comparator. As shown in Fig. \ref{fig_signal_folding}(a), a 1-bit capacitive DAC is used to implement the signal folding circuit in this design. Fig. \ref{fig_signal_folding}(b) presents the timing diagram of the circuit. When it receives a clock signal produced by the refractory circuit, it will switch on the connection with $V_{M}$, and thus fold the input signal to the value of $V_{M}$ by charge redistribution. Then, the switch turns off, and the signal folding circuit continues to track the input signal. Two switches are used to deal with the asynchronous clock signals from up-crossing and down-crossing. Due to the fixed $V_{M}$ connected with switches, charge injection is independent of input signals and converted to offsets that could be eliminated by proper calibration.

\subsection{Dynamic Comparator with Static Power Reduction}
\begin{figure}[t]
	\centerline{\includegraphics[width=3in]{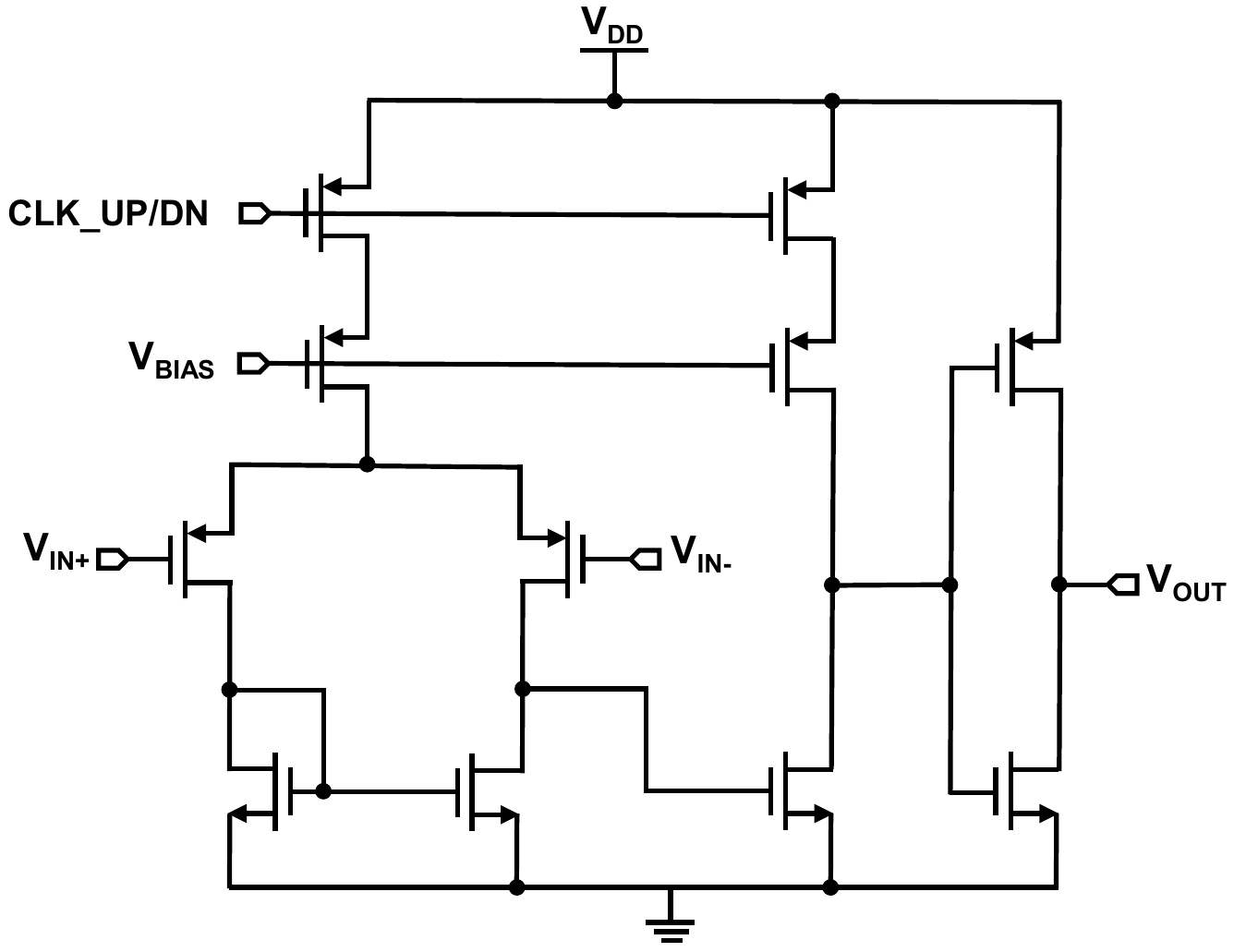}}
	\vspace{-0.3cm}
	\caption{Schematic of the dynamic comparator.}
	\vspace{-0.4cm}
	\label{fig_comparator}
\end{figure}

Comparators are designed for comparing input signal with reference voltage levels to get digital decision output. Two dynamic comparators are used in Neuron-ADC to compare input signal with a high reference voltage level and a low reference voltage level, respectively. A 3-stage dynamic comparator is utilized and presented in Fig. \ref{fig_comparator}. A differential amplifier with PMOS input is exploited as the first stage, followed by one common source stage to increase gain performance and reduce the kickback noise. To achieve a rail-to-rail swing digital output, one inverter stage is employed at the output. Additionally, two power-gating PMOS transistors are added and controlled by the refractory circuit to turn off comparators during the signal folding process, which helps to reduce static power consumption.

\subsection{Bio-Inspired Refractory Circuit}
\begin{figure}[t]
	\centerline{\includegraphics[width=3.5in]{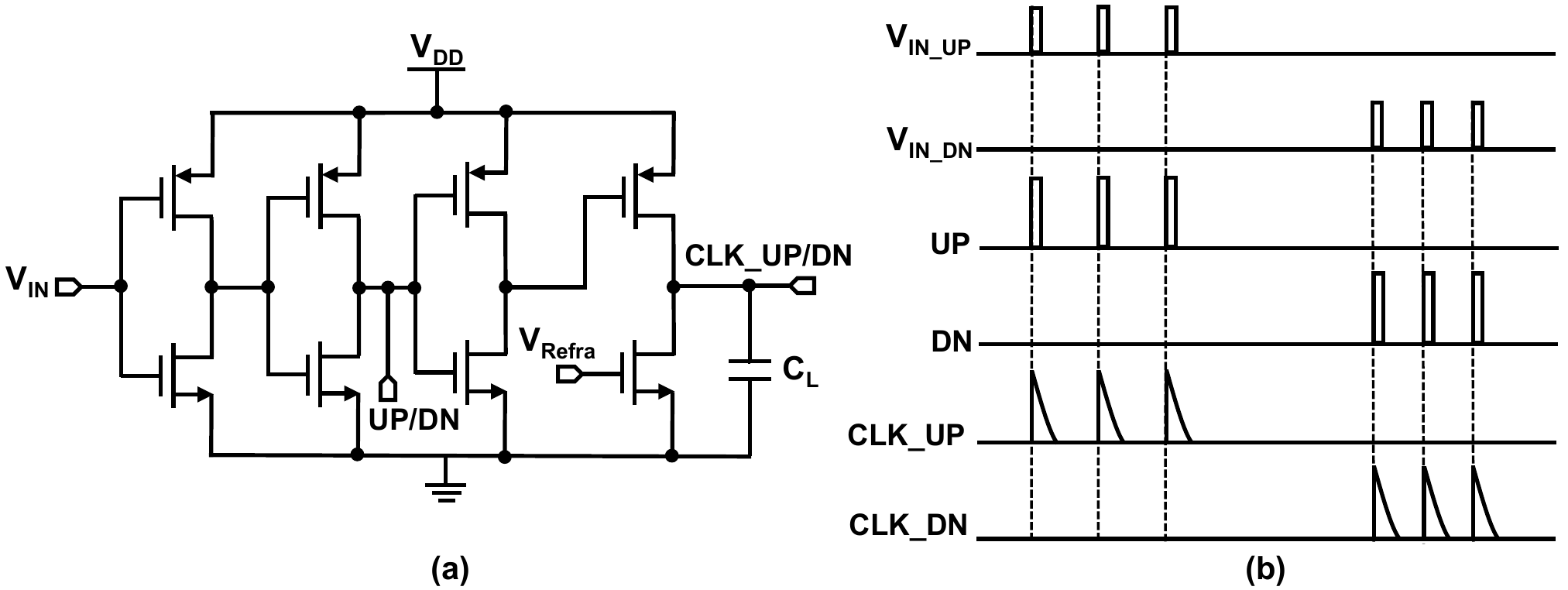}}
	\vspace{-0.3cm}
	\caption{The bio-inspired refractory circuit. (a) Circuit schematic. (b) Timing diagram.}
	\vspace{-0.3cm}
	\label{fig_refractory}
\end{figure}

The bio-inspired refractory circuit is proposed to achieve reconfigurable sampling of Neuron-ADC. The refractory circuit is inspired by biological neurons and prior circuit implementations \cite{sharifshazileh2021electronic}. This is the first time the refractory circuit and level-crossing sampling are combined to achieve reconfigurable and compressive sampling. Fig. \ref{fig_refractory} shows the schematic and timing diagram of the bio-inspired refractory circuit. It includes a PMOS common source stage with an adjustable NMOS load connected with inverter buffers. By changing the refractory voltage of the NMOS load, the refractory circuit generates asynchronous clock signals with variable refractory period shown as CLK\_UP/CLK\_DN in Fig. \ref{fig_refractory}(b). Low refractory voltage produces a long refractory period for the signal folding circuit, leading to a long signal folding time and reduced number of sampled data points. High refractory voltage works vice versa. Reconfigurable sampling is thus realized by changing the refractory voltage.

\section{Simulation Results}

The proposed Neuron-ADC is simulated using TSMC 40 nm CMOS process to evaluate its performance. The supply voltage of all circuit blocks is set to 0.6 V to reduce power consumption. The value of LSB is closely related to the power and operation condition of Neuron-ADC. LSB is calculated by 
\begin{equation}
LSB=\frac{A_{FS}}{2^{M}}
\label{equ:LCADC}
\end{equation}
where $A_{FS}$ is the full-scale voltage range of input signal. M is the resolution bit of Neuron-ADC, producing ${2^{M}}$ quantization levels. To balance the trade-off between power consumption and signal conversion linearity, $V_H$, $V_M$ and $V_L$ are set to 95 mV, 75 mV, and 55mV, respectively. Therefore, 1 LSB is 20 mV. A sinusoidal signal ranging from 10 Hz to 10 kHz with 640 mV amplitude is used as the input. Fifth-order polynomial interpolation is performed in MATLAB to calculate the FFT and signal-to-noise-and-distortion ratio (SNDR) of the simulation results.

\begin{figure}[t]
	\centerline{\includegraphics[width=3.3in]{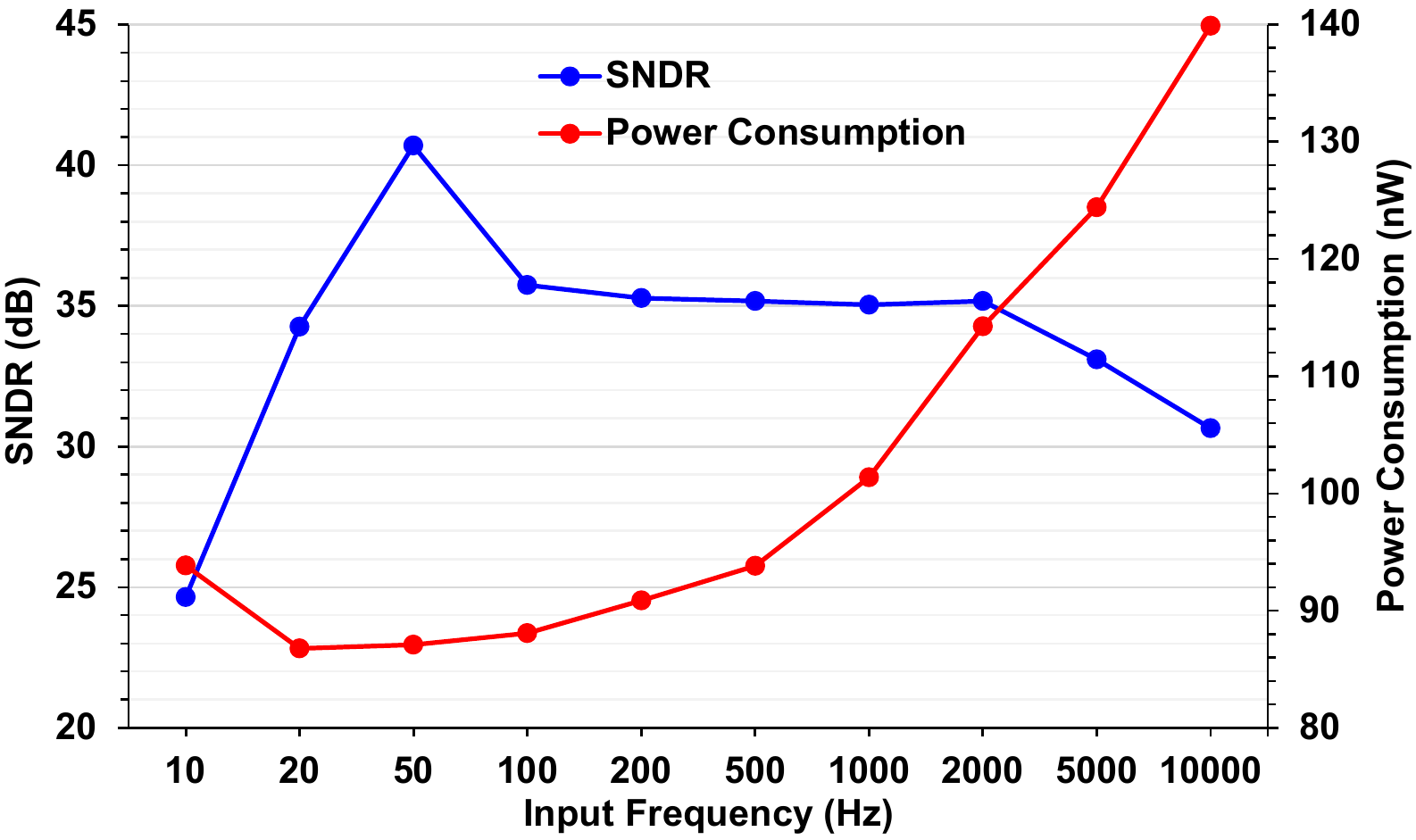}}
	\vspace{-0.3cm}
	\caption{SNDR and power consumption as a function of input frequency from 10 Hz to 10 kHz. Refractory voltage is 200 mV for each frequency point.}
	\vspace{-0.6cm}
	\label{fig_SNDR_power}
\end{figure}

Fig. \ref{fig_SNDR_power} presents the SNDR and power consumption as a function of input frequency from 10 Hz to 10 kHz with 640 mV amplitude. To have a fair comparison, the refractory voltage is 200 mV for each frequency point. As shown in Fig. \ref{fig_SNDR_power}, a higher input frequency typically leads to more power consumption because of high dynamic power. From 10 Hz to 50 Hz, the SNDR increases and peaks at 50 Hz input frequency, after which it decreases. For low input frequency, the SNDR is affected by the operation accuracy of the signal folding circuit and comparators. While for high input frequency, the overall loop delay of the Neuron-ADC limits the SNDR. It is worth mentioning that different refractory voltage results in different power consumption and SNDR given the same input frequency.

\begin{figure}[t]
	\centerline{\includegraphics[width=3.3in]{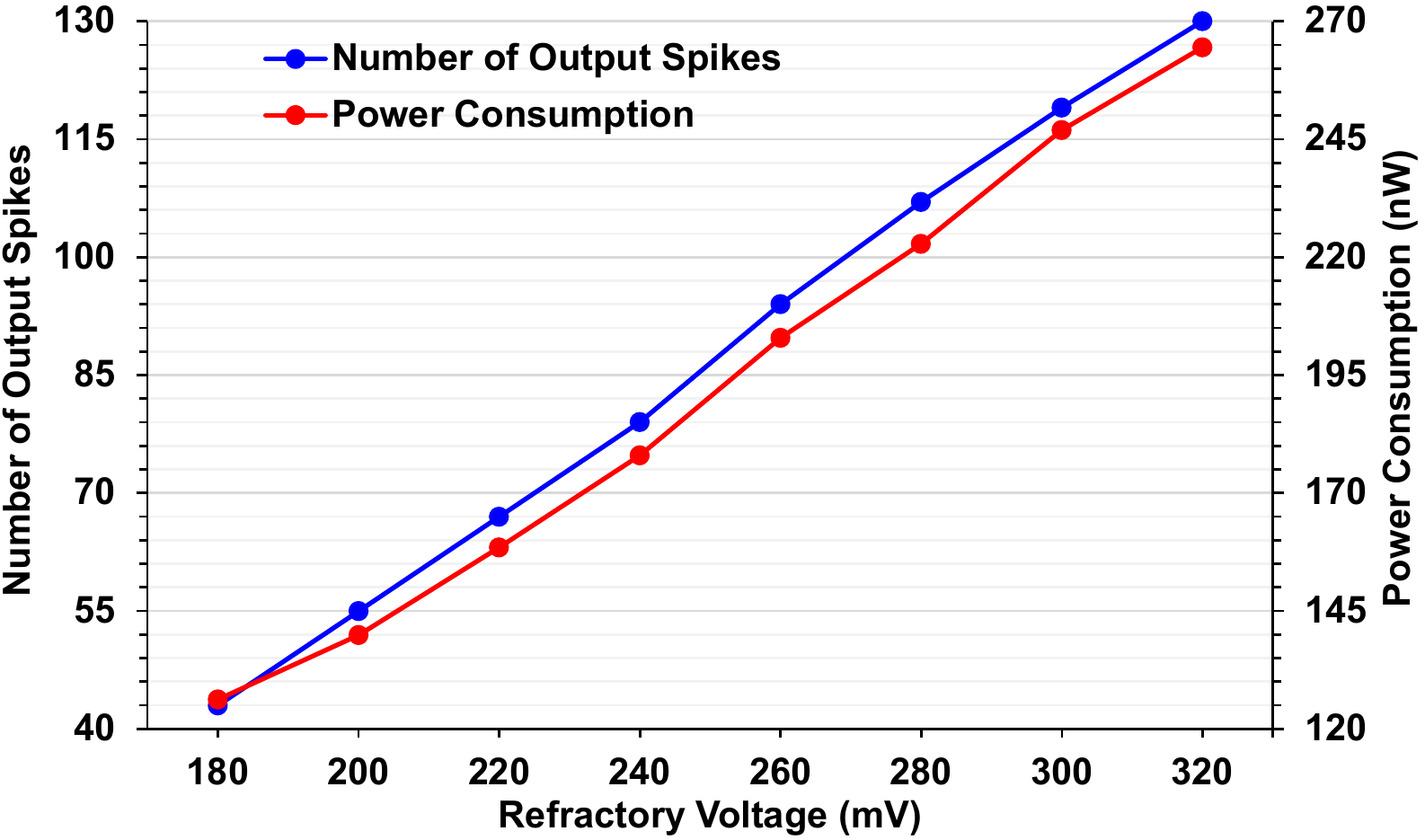}}
	\vspace{-0.3cm}
	\caption{Number of output spikes and power consumption as a function of refractory voltage given 10 kHz sinusoidal input.}
	\vspace{-0.5cm}
	\label{fig_power_spike}
\end{figure}

To demonstrate the reconfigurable sampling function of the Neuron-ADC, Fig. \ref{fig_power_spike} illustrates the power consumption and number of output spikes as a function of refractory voltage given 10 kHz sinusoidal input. As illustrated in Fig. \ref{fig_power_spike}, the power consumption and number of output spikes both increase with refractory voltage. Reconfigurable sampling can be achieved by adjusting the value of refractory voltage, which offers a flexible option for circuit designers. If the major design target is low power consumption, a low refractory voltage is required. While for high conversion accuracy, a high refractory voltage is more suitable. Moreover, the maximum output rate of the Neuron-ADC is reconfigurable for various bio-signals with different waveform shapes and slope features.

\begin{table}[t]
	\centering
	\label{table_static_power}
	\caption{Static power reduction of comparators.}
	\vspace{-0.1cm}
	\begin{tabular}{|c|c|c|c|} 
		\hline
		\multirow{2}{*}{\begin{tabular}[c]{@{}c@{}}Input\\Frequency (Hz)\end{tabular}} & \multicolumn{2}{c|}{Static Power (nW)$^{\mathrm{*}}$}  & \multirow{2}{*}{\begin{tabular}[c]{@{}c@{}}Reduction\\ Percent\end{tabular}}  \\ 
		\cline{2-3}
		& W/O Reduction         & With Reduction &                                                                               \\ 
		\hline
		500                                                                           & \multirow{5}{*}{30.7} & 26.1         & 14.8\%                                                                       \\ 
		\cline{1-1}\cline{3-4}
		1000                                                                          &                       & 24.2         & 21.1\%                                                                       \\ 
		\cline{1-1}\cline{3-4}
		2000                                                                          &                       & 22.5         & 26.6\%                                                                       \\ 
		\cline{1-1}\cline{3-4}
		5000                                                                          &                       & 19.6         & 36.0\%                                                                       \\ 
		\cline{1-1}\cline{3-4}
		10000                                                                         &                       & 18.1         & 41.1\%                                                                       \\
		\hline
		\multicolumn{4}{l}{$^{\mathrm{*}}$Refractory voltage is 100 mV.}
	\end{tabular}
	\vspace{-0.5cm}
\end{table}

\begin{table}[t!]
	\label{table_comparison}
	\caption{Performance comparison of state-of-the-art works.}
	\vspace{-0.1cm}
	\resizebox{\linewidth}{!}{%
		\begin{tabular}{|c|c|c|c|c|c|} 
			\hline
			Parameter                                                        & \cite{hou20181} & \begin{tabular}[c]{@{}c@{}}\cite{yazdani2021low}\\(Simulation)\end{tabular} & \cite{marisa2017pseudo} & \cite{hou201861} & \begin{tabular}[c]{@{}c@{}}This work\\(Simulation)\end{tabular}  \\ 
			\hline
			Technology (nm)                                                  & 180                              & 180                                                                                          & 350                                      & 180                               & 40                                                               \\ 
			\hline
			Supply (V)                                                       & 0.55-1                           & 0.8                                                                                          & 1.8-2.4                                  & 0.5                               & 0.6                                                              \\ 
			\hline
			Bandwidth (kHz)                                                  & 1                                & 4                                                                                            & 1                                        & 1                                 & 10                                                               \\ 
			\hline
			SNDR (dB)                                                        & 39-49                            & 38.5-43                                                                                      & 37-48                                    & 35                                & 43.2                                                             \\ 
			\hline
			ENOB (bit)                                                       & 6.2-7.9                          & 6.1-6.8                                                                                      & 4-8                                      & 5.6                               & 6.9                                                              \\ 
			\hline
			Power (nW)                                                       & 4.2-186                          & 90-180                                                                                       & 600-2000                                 & 60-220                            & 229.8                                                            \\ 
			\hline
			FoM (fJ/conv.)                                                   & 1400                             & 103-343                                                                                      & 1140                                     & 627                               & 97                                                               \\ 
			\hline
			\begin{tabular}[c]{@{}c@{}}Reconfigurable\\Sampling\end{tabular} & No                               & No                                                                                           & No                                       & Yes                               & Yes                                                              \\ 
			\hline
			\begin{tabular}[c]{@{}c@{}}Static Power \\Reduction\end{tabular} & No                               & Yes                                                                                          & No                                       & No                                & Yes                                                              \\
			\hline
		\end{tabular}
	}
	\vspace{-0.7cm}
\end{table}

The above simulation results are without the function of static power reduction. Table 1 presents the simulation results of static power reduction of comparators with different input frequency. The refractory voltage is related to power reduction and is set to 100 mV. From 500 Hz to 10 kHz input, the static power reduction percent increases from 14.8\% to 41.1\%. Given the same refractory voltage, a higher input frequency produces more output spikes and longer refractory period, making comparators turn off longer and saving more static power.

Simulation results show that the proposed Neuron-ADC can achieve a FoM of 97 fJ/conversion given input signal frequency at 10 kHz and refractory voltage of 285 mV without static power reduction function. The FoM is calculated by equation (\ref{equ:FoM}). Table 2 compares the Neuron-ADC with state-of-the-art works. This work achieves the lowest FoM and realizes both reconfigurable sampling and static power reduction.
\vspace{-0.3cm}
\begin{equation}
FoM=\frac{Power}{2BW\cdot2^{ENOB}}
\label{equ:FoM}
\end{equation}

\section{Conclusion}
This work proposes a bio-inspired Neuron-ADC with reconfigurable sampling and static power reduction for biomedical applications. The Neuron-ADC utilizes level-crossing compressive sampling and converts bio-signal to digital spikes to prevent wasting ADC energy on unnecessary data. Three design challenges of Neuron-ADC have been addressed: charge-injection-free signal folding circuit, dynamic comparator with static power reduction, and bio-inspired refractory circuit. The reconfigurable sampling function makes the Neuron-ADC suitable for various bio-signals with either low power operation or high conversion accuracy. Moreover, the dynamic comparator can reduce static power up to 41.1\% with 10 kHz sinusoidal test input. Simulation results of 40 nm CMOS technology show that from 10 Hz to 10 kHz input bandwidth, the ENOB is up to 6.9 bits with a corresponding FoM of 97 fJ/conversion. To sum up, the proposed Neuron-ADC paves the way to low-power and reconfigurable multi-modal biomedical recording and neuromorphic processing systems.

\section*{Acknowledgment}
The authors would like to acknowledge funding support from Westlake University, Zhejiang Leading Innovative and Entrepreneur Team Introduction Program No. 2020R01005, and Zhejiang Key R\&D Program No. 2021C03002.

{\small{
		\bibliographystyle{IEEEtran}
		\bibliography{reference.bib}
		}
}

\end{document}